\documentclass[letter,longauth,traditabstract]{aa}

\usepackage{graphicx}

\usepackage{txfonts}

\begin{document}

\title{The \textit{Herschel}-SPIRE instrument and its in-flight performance
\thanks{\textit{Herschel} is an ESA space observatory with science instruments provided by European-led Principal Investigator consortia and with important participation from NASA.}
}

\author{M. J. Griffin\inst{1}
\and A. Abergel \inst{2} 
\and A. Abreu \inst{3} 
\and P. A. R. Ade \inst{1} 
\and P. Andr\'{e} \inst{4} 
\and J.-L. Augueres \inst{4} 
\and T. Babbedge \inst{5} 
\and Y. Bae \inst{6} 
\and T. Baillie \inst{7} 
\and J.-P. Baluteau \inst{8}
% ...
}

\institute{School of Physics and Astronomy, Cardiff University, The Parade, Cardiff, CF24 3AA, UK
\and Institut d'Astrophysique Spatiale, Universit\'{e} Paris-Sud 11, 91405 Orsay, France
\and ESA Herschel Science Centre, ESAC, Villaneuva de la Ca\~{n}ada, Spain
\and Commissariat \`{a} l'\'{E}nergie Atomique, Service d'Astrophysique, Saclay, 91191 Gif-sur-Yvette, France
\and Imperial College London, Blackett Laboratory, Prince Consort Road, London SW7, 2AZ, UK
\and NASA Jet Propulsion Laboratory, 4800 Oak Grove Drive, Pasadena, CA 91109, USA
\and UK Astronomy Technology Centre, Royal Observatory, Blackford Hill, Edinburgh EH9 3HJ, UK
\and Laboratoire d'Astrophysique de Marseille, UMR6110 CNRS, 38 rue F. Joliot-Curie, F-13388 Marseille, France
% ...
}

\date{Received March 31 2010; accepted April 21 2010}


\abstract
{The Spectral and Photometric Imaging Receiver (SPIRE), is the \textit{Herschel} Space Observatory`s submillimetre camera and spectrometer.  It contains a three-band imaging photometer operating at 250, 350 and 500 $\mu$m, and an imaging Fourier Transform Spectrometer (FTS) which covers simultaneously its whole operating range of 194-671 $\mu$m (447-1550 GHz).  The SPIRE detectors are arrays of feedhorn-coupled bolometers cooled to 0.3 K.  The photometer has a field of view of 4\arcmin\ x 8\arcmin, observed simultaneously in the three spectral bands.  Its main operating mode is scan-mapping, whereby the field of view is scanned across the sky to achieve full spatial sampling and to cover large areas if desired.  The spectrometer has an approximately circular field of view with a diameter of 2.6 \arcmin. The spectral resolution can be adjusted between 1.2 and 25 GHz by changing the stroke length of the FTS scan mirror. Its main operating mode involves a fixed telescope pointing with multiple scans of the FTS mirror to acquire spectral data.  For extended source measurements, multiple position offsets are implemented by means of an internal beam steering mirror to achieve the desired spatial sampling and by rastering of the telescope pointing to map areas larger than the field of view.  The SPIRE instrument consists of a cold focal plane unit located inside the \textit{Herschel} cryostat and warm electronics units, located on the spacecraft Service Module, for instrument control and data handling.  Science data are transmitted to Earth with no on-board data compression, and processed by automatic pipelines to produce calibrated science products.   The in-flight performance of the instrument matches or exceeds predictions based on pre-launch testing and modelling: the photometer sensitivity is comparable to or slightly better than estimated pre-launch, and the spectrometer sensitivity is also better by a factor of 1.5 –- 2.}



\maketitle

\section{Introduction}

The SPIRE instrument is designed to exploit the particular advantages of the \textit{Herschel} Space Observatory (Pilbratt et al. 2010) for observations in the submillimetre region: its large (3.5-m), cold ($\sim$ 85-K), low-emissivity ($\sim$ 1\%) telescope; unrestricted access to the poorly explored 200-700 $\mu$m range; and the large amount (over 20,000 hrs) of observing time. In this paper we summarise the key design features of the instrument, outline its main observing modes, and present a summary of its measured in-flight performance and scientific capabilities.  Further details of the instrument calibration are given in Swinyard et al. (2010).

\section{SPIRE instrument design}
SPIRE consists of a three-band imaging photometer and an imaging Fourier Transform Spectrometer (FTS). The photometer carries out broad-band photometry ($\lambda$/$\Delta\lambda$ $\sim$ 3) in three spectral bands centred on approximately 250, 350 and 500 $\mu$m, and the FTS uses two overlapping bands to cover 194-671 $\mu$m (447-1550 GHz).  

Figure 1 shows a block diagram of the instrument. The SPIRE focal plane unit (FPU) is approximately 700 x 400 x 400 mm in size and is supported from the 10-K \textit{Herschel} optical bench by thermally insulating mounts. It contains the optics, detector arrays (three for the photometer, and two for the spectrometer), an internal $^{3}$He cooler to provide the required detector operating temperature of $\sim$ 0.3 K, filters, mechanisms, internal calibrators, and housekeeping thermometers.  It has three temperature stages: the \textit{Herschel} cryostat provides temperatures of 4.5 K  and 1.7 K via high thermal conductance straps to the instrument, and the $^{3}$He cooler serves all five detector arrays. 

The photometer and FTS both have cold pupil stops conjugate with the \textit{Herschel} secondary mirror, which is the telescope system pupil, defining a 3.29-m diameter used portion of the primary.  Conical feedhorns (Chattopadhyay et al. 2003) provide a roughly Gaussian illumination of the pupil, with an edge taper of around 8 dB in the case of the photometer. The same $^{3}$He cooler design (Duband et al. 2008) is used in SPIRE and in the PACS instrument (Poglitsch et al. 2010). It has two heater-controlled gas gap heat switches; thus one of its main features is the absence of any moving parts. Liquid confinement in zero-$g$ is achieved by a porous material that holds the liquid by capillary attraction.  A Kevlar wire suspension system supports the cooler during launch whilst minimising the parasitic heat load. The cooler contains 6 STP litres of $^{3}$He, fits in a 200 x 100 x 100-mm envelope and has a mass of $\sim$ 1.7 kg. Copper straps connect the 0.3-K stage to the five detector arrays, and are held rigidly at various points by Kevlar support modules (Hargrave et al. 2006). The supports at the entries to the 1.7-K boxes are also light-tight.

All five detector arrays use hexagonally close-packed feedhorn-coupled spider-web Neutron Transmutation Doped (NTD) bolometers (Turner et al. 2001).  The bolometers are AC-biased with frequency adjustable between 50 and 200 Hz, avoiding 1/\textit{f} noise from the cold JFET readout readouts. There are three SPIRE warm electronics units: the Detector Control Unit (DCU) provides the bias and signal conditioning for the arrays and cold electronics, and demodulates and digitises the detector signals; the FPU Control Unit (FCU) controls the cooler and the mechanisms, and reads out all the FPU thermometers; and the Digital Processing Unit (DPU) runs the on-board software and interfaces with the spacecraft for commanding and telemetry.  
\begin{figure}
\begin{center}
\includegraphics[width=6.0cm,angle=-90]{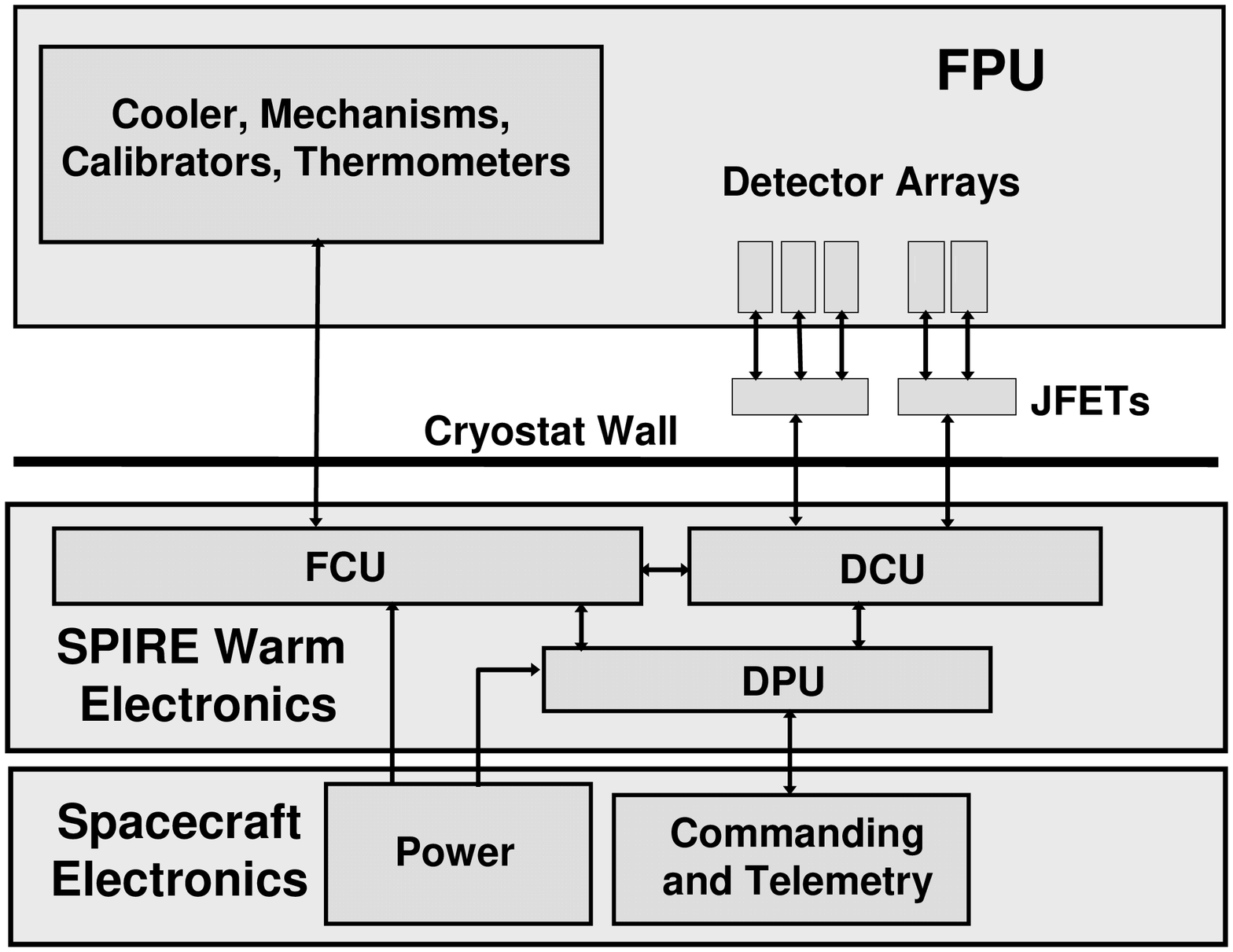}
\caption{SPIRE instrument architecture}
\end{center}
\end{figure}
\section{Photometer design}
Figure 2 shows the opto-mechanical layout of the photometer side of the FPU.  It is an all-reflective design (Dohlen et al. 2000) except for the dichroics used to direct the three bands onto the bolometer arrays, and the filters used to define the passbands (Ade et al. 2006). The image is diffraction-limited over the 4\arcmin\  x 8\arcmin\ field of view, which is offset by 11\arcmin\  from the centre of the \textit{Herschel} telescope's highly curved focal surface.  The input mirror M3, lying below the telescope focus, receives the \textit{f}/8.7 telescope beam and forms an image of the secondary at the flat beam steering mirror (BSM), M4. A calibration source (Pisano et al. 2005) placed behind a hole in the centre of the BSM, is used to provide a repeatable signal for the bolometers. It occupies an area contained within the region of the pupil obscured by the hole in the primary. Mirror M5 converts the focal ratio to \textit{f}/5 and provides an intermediate focus at M6, which re-images the M4 pupil to a cold stop. The input optics are common to the photometer and spectrometer and the separate spectrometer field of view is directed to the other side of the optical bench panel by a pick-off mirror close to M6.  The 4.5-K optics are mounted on the SPIRE internal optical bench.  Mirrors M7, M8 and a subsequent mirror inside the 1.7-K box form a one-to-one optical relay to bring the M6 focal plane to the detectors. The 1.7-K enclosure also contains the three detector arrays, and two dichroic beam splitters to direct the same field of view onto the arrays so that it can be observed simultaneously in the three bands.  
\begin{figure}
\begin{center}
\includegraphics[width=7.0cm,angle=-90]{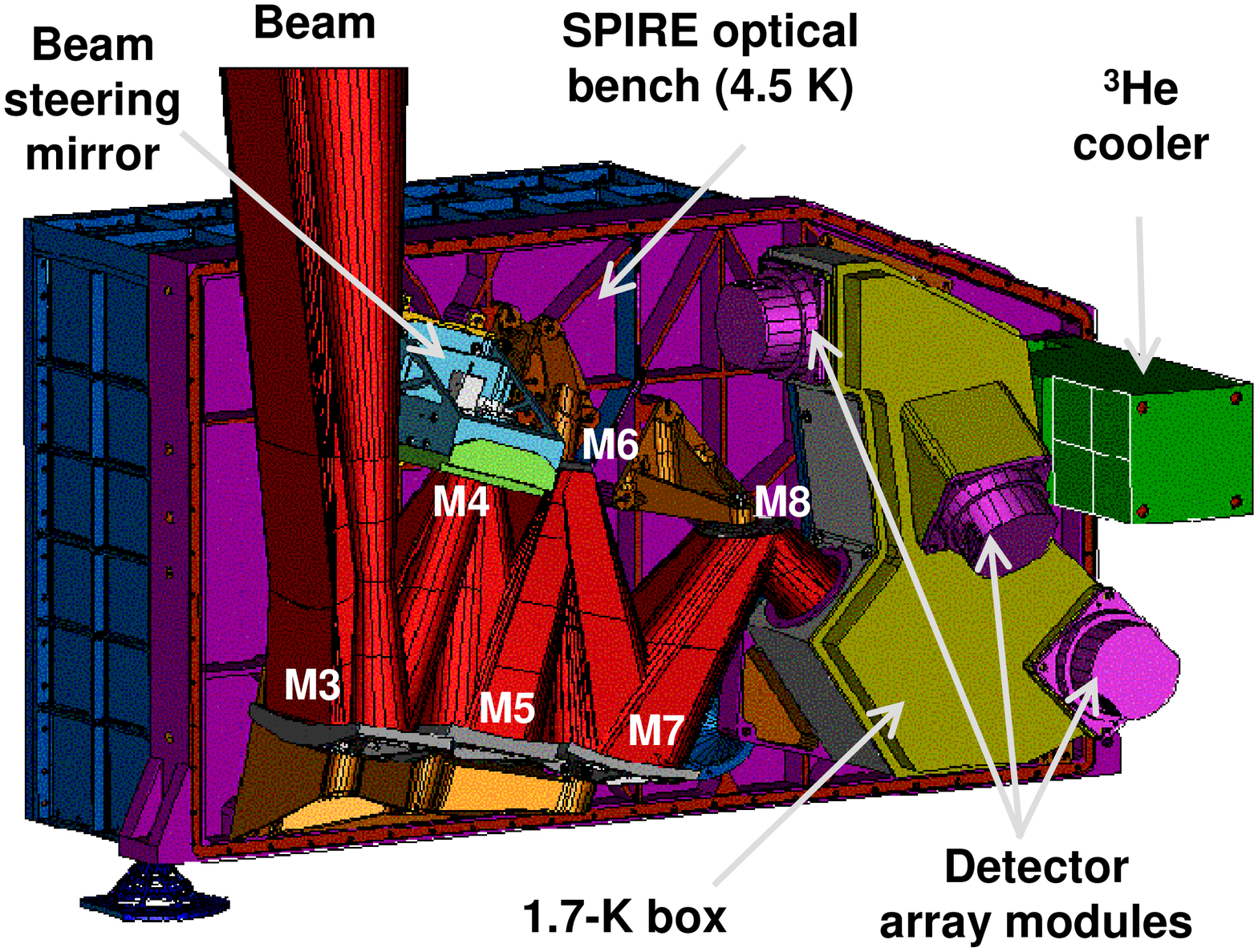}
\caption{SPIRE FPU: photometer side layout}
\end{center}
\end{figure}

The bolometer array modules are bolted to the outside wall of the 1.7-K box.  Inside each one, the $^{3}$He stage, accommodating the detectors, feedhorns and final filters, is thermally isolated from the 1.7-K mount by tensioned Kevlar threads, and cooled by a thermal strap to the $^{3}$He cooler.   The three arrays contain 43 (500 $\mu$m), 88 (350 $\mu$m) and 139 (250 $\mu$m) detectors. The relative merits of feedhorn-coupled detectors, as used by SPIRE, and filled array detectors, which are used by \textit{Herschel}-PACS (Poglitsch et al. 2010) and some ground-based instruments such as SCUBA-2 (Audley et al. 2007) and SHARC-II (Dowell et al. 2003), are discussed in detail in Griffin et al. (2002).  In the case of SPIRE, the feedhorn-coupled architecture was chosen as the best option given the achievable sensitivity, the requirements for the largest possible field of view and high stray light rejection, and limitations on the number of detectors imposed by spacecraft resource budgets.  The detector feedhorns are designed for maximum aperture efficiency, requiring an entrance aperture close to 2$F\lambda$, where $\lambda$ is the wavelength and $F$ is the final optics focal ratio.  This corresponds to a beam spacing on the sky of 2$\lambda/D$, where $D$ is the telescope diameter.  The array layouts are shown schematically in Fig. 3 which also shows a  photograph of an array module. 
\begin{figure}
\begin{center}
\includegraphics[width=8.0cm,angle=0]{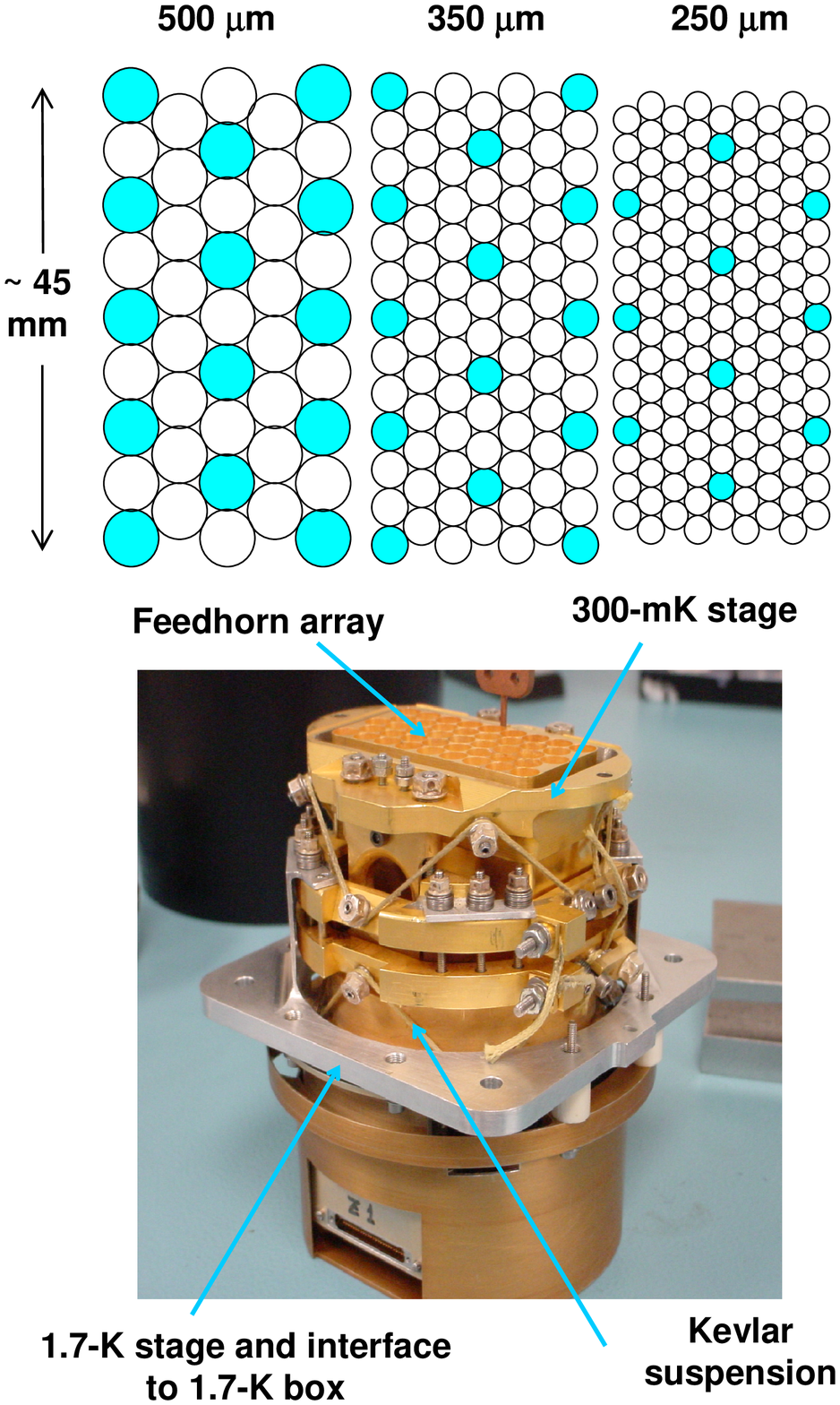}
\caption{Top: layout of the photometer arrays with shaded circles representing detectors for which there is overlap on the sky for the three bands; bottom: photograph of a SPIRE detector array module}
\end{center}
\end{figure}

The photometric passbands are defined by quasi-optical edge filters (Ade et al. 2006) located at the instrument input, at the 1.7-K cold stop, and directly in front of the detector arrays, the reflection-transmission edges of the dichroics, and the cut-off wavelengths of the feedhorn output waveguides. The filters also serve to minimise the thermal loads on the 1.7-K and 0.3-K stages. 

The AC-biased bolometer signals are de-modulated by individual lock-in amplifiers (LIAs) in the DCU.  Each LIA comprises a bandpass filter and a square wave demodulator, followed by a 5-Hz low-pass filter.  The output of the LIA is a DC voltage proportional to the RMS value of the voltage at the bolometer output. The LIA outputs are multiplexed and sampled at nominally 18.6 Hz for telemetry to the ground.  In order to achieve the necessary 20-bit sampling using a 16-bit analogue-to-digital converter, an offset subtraction scheme is implemented.  After multiplexing, a suitable 4-bit DC offset (generated on board and transmitted to the ground in the science data stream) is subtracted from each signal prior to the final gain stage before digitisation. The 130 kbs available data rate allows the data to be transmitted to the ground with no on-board processing.

\section{Spectrometer design}
The FTS (Swinyard et al. 2003; Dohlen et al. 2000) uses two broadband intensity beam splitters in a Mach-Zehnder configuration which has spatially separated input and output ports.  One input port views a 2.6\arcmin\  diameter field of view on the sky and the other an on-board reference source.  Two bolometer arrays at the output ports cover overlapping bands of 194-313 $\mu$m (SSW) and 303-671 $\mu$m (SLW). As with any FTS, each scan of the moving mirror produces an interferogram in which the spectrum of the entire band is encoded with the spectral resolution corresponding to the maximum mirror travel.

The FTS focal plane layout is shown in Fig. 4. A single back-to-back scanning roof-top mirror serves both interferometer arms. It has a frictionless mechanism using double parallelogram linkage and flex pivots, and a Moir\'{e} fringe sensing system.  A filtering scheme similar to the one used in the photometer restricts the passbands of the detector arrays at the two ports, defining the two overlapping FTS bands.  
\begin{figure}
\begin{center}
\includegraphics[width=7.0cm,angle=-90]{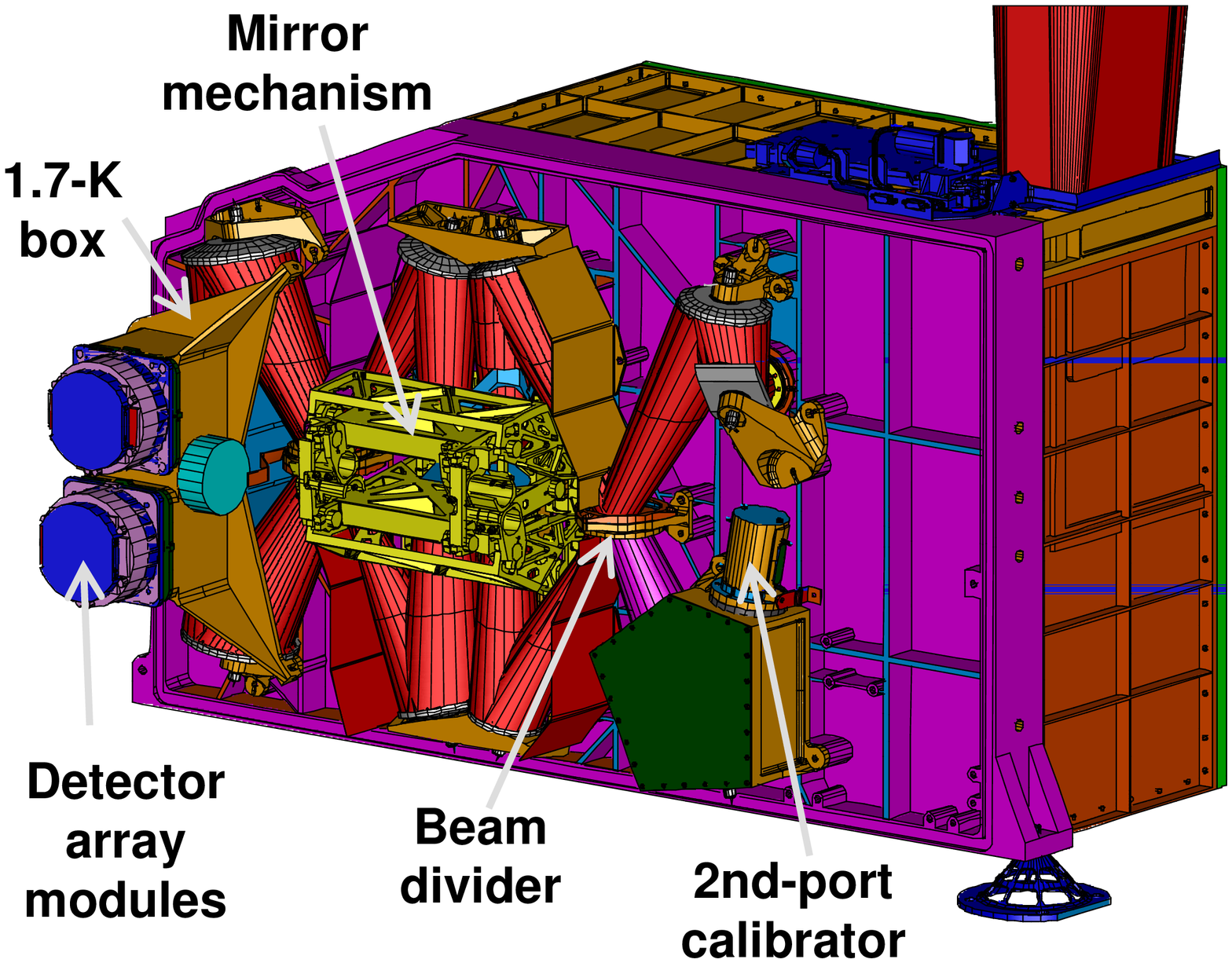}
\caption{SPIRE FPU: spectrometer side layout}
\end{center}
\end{figure}

The FTS spectral resolution element is given by 1/(2$d$) where $d$ is the maximum optical path difference. The highest unapodised resolution available is 0.04 cm$^{-1}$ (1.2 GHz), corresponding to a FWHM of the instrument spectral response function of 0.048 cm$^{-1}$.  For this resolution, $\lambda$/$\Delta$$\lambda$ varies between $\sim$ 1300 at the short-wavelength end and 370 at the long-wavelength end.  The standard unapodised resolution adopted for spectrophotometry is 0.83 cm$^{-1}$ for which $\lambda$/$\Delta$$\lambda$ varies from approximately 20 to 60 between the long- and short-wavelength ends of the range.

The two hexagonally close-packed spectrometer arrays contain 37 detectors in the short-wavelength array and 19 in the long-wavelength array.  The array modules are similar to those used for the photometer, with an identical interface to the 1.7-K enclosure.  The feedhorn and detector cavity designs are optimised to provide good sensitivity across the whole wavelength range of the FTS.  The SSW feedhorns are sized to give 2$F\lambda$ pixels at 225 $\mu$m and the SLW horns are 2$F\lambda$ at 389 $\mu$m.  This arrangement has the advantage that there are many co-aligned pixels in the combined field of view.  The SSW beams on the sky are 33\arcsec\  apart, and the SLW beams are separated by 51\arcsec. A thermal source, SCal (Hargrave et al. 2003), at the second input port is available to allow the background power from the telescope to be nulled, thereby reducing the dynamic range needed in recording the interferograms (the amplitude of the interferogram central maximum is proportional to the difference in the radiant power from the two ports). 

The FTS on-board signal chain is similar to that for the photometer, but has a higher low-pass-filter cut-off frequency of 25 Hz to accommodate the required signal frequency range, and the sampling rate is correspondingly higher (80 Hz).  As with the photometer, the 130 kbs available data rate is sufficient to allow data transmission to Earth with no on-board processing.

\section{Observing modes}
The photometer was designed to have three principal observing modes: point source photometry, field (jiggle) mapping, and scan mapping.  The first two modes involve chopping and jiggling with the SPIRE beam steering mirror and nodding using the telescope.  In practice, because of the excellent performance and simplicity of the scan-map mode, it is the optimum for most observations.  Although its efficiency for point source and small map observations is low, it provides better data quality, and over a larger area, than the chopped modes, and can produce a sky background confusion limited map in a total observing time that is still dominated by telescope slewing overheads. 

In scan-map mode, the telescope is scanned across the sky at 30 or 60\arcsec/s. The scan angle is chosen to give the beam overlap needed for full spatial sampling over a strip defined by one scan line, and to provide a uniform distribution of integration time over the area covered by the scan. One map repeat is normally constituted by two scans in orthogonal directions to provide additional data redundancy and cross-linking. As well as SPIRE-only and PACS-only scan-map modes, \textit{Herschel} can implement simultaneous five-band photometric scan mapping in SPIRE-PACS Parallel mode.  Data are taken simultaneously in the three SPIRE bands and two of the PACS bands, providing a highly efficient means of making multi-band maps of large areas (at least 30\arcmin\ x 30\arcmin).  Scan speeds of 20 or 60\arcsec/s are available in Parallel mode. The SPIRE detector sampling rate in this mode is reduced to 10 Hz to keep within the data rate budget, with minimal impact on the data quality.

The standard operating mode for the FTS is to scan the mirror at constant speed (nominally 0.5 mm/s) with the telescope pointing fixed, providing an optical path rate of change of 2 mm/s due to the factor of four folding in the optics.  Radiation frequencies of interest are encoded by the scanning motion as detector output electrical frequencies in the range 3-10 Hz.  The maximum scan length is 3.5 cm, corresponding to an optical path difference of 14 cm.  Three values of spectral resolution are adopted as standard: high, medium, and low, giving unapodised resolutions of 0.04, 0.24, and 0.83 cm$^{-1}$ respectively (corresponding to 1.2, 7.2, and 25 GHz).

For point source observations with the FTS, the source is positioned at the overlapping SSW and SLW detectors at the array centres; but data are acquired for all of the detectors, providing at the same time a sparsely-sampled map of the emission from the region around the source.  Likewise, a single pointing generates a sparse map of an extended object.  Full and intermediate spatial sampling can be obtained using the SPIRE BSM to provide the necessary pointing changes between scans while the telescope pointing remains fixed. Regions larger than the field of view are mapped using telescope rastering combined with spatial sampling with the BSM.  Full details of the SPIRE observing modes are given in the SPIRE Observers' Manual (2010).

\section{Instrument in-flight optimisation and achieved performance}
\subsection{In-flight functional performance}
All SPIRE subsystems are fully functional and meet their key requirements.  The $^{3}$He cooler has a hold-time in excess of 46 hrs, allowing two full days of SPIRE operation after a cooler cycle.  The achieved temperatures are $\sim$ 287 mK at the $^{3}$He cold tip, 310 mK for the photometer arrays and 315 mK for the FTS arrays.  These values are in line with on-ground measurements and slightly better than used in the pre-launch instrument sensitivity model.  Detector yield is excellent, with totals of only six unusable photometer bolometers and two unusable FTS bolometers. In-flight testing, carried out before the \textit{Herschel} cryostat lid was removed, allowed the main results of system-level pre-flight testing (with \textit{Herschel} in the Large Space Simulator facility at ESTEC) to be reproduced.  The radiant background on the detectors, from the thermal emission of the \textit{Herschel} telescope and any additional stray light, is consistent with on-ground characterisation of the \textit{Herschel} telescope emissivity (Fischer et al. 2004), with a stray light level somewhat less than that adopted in pre-launch estimation of the  performance.   It is clear that the launch of \textit{Herschel} proved to be a rather benign event for the system.

\subsection{Instrument optimisation and AOT characterisation}
SPIRE has many parameters which are adjustable in flight to achieve best performance.  These include: cooler recycle settings and detailed timing; JFET pre-amplifier supply voltage settings; detector array bias frequencies and voltages; detector lock-in amplifier phase settings; PCal and SCal applied power settings; telescope angular scan speeds and scan angles for scan-map observations; FTS mirror and BSM servo system parameters; and the FTS mirror scan speed. A good part of the Commissioning and Performance Verification (PV) phases of the \textit{Herschel} mission involved systematic measurements and analysis to optimise the values of these parameters. In the case of the detector-related settings, separate optimisations were done for nominal source strength ($<$ 200 Jy for the photometer) and bright source settings.  Most parameter optimisations resulted in values close to or identical to those predicted before launch.  Important differences were: (i) the necessary frequency of PCal operation to track detector responsivity, planned for at least once per hour pre-launch, has been relaxed to once per observation regardless of the observation length; (ii) because of the low telescope and stray light background, it is possible to operate the FTS without switching on the SCal unit.  This has the advantages that the operation and data reduction are simplified, and that the absence of SCal thermal emission reduces photon noise.

\subsection{Glitches due to ionising radiation}
In space, ionising radiation hits deposit thermal energy in the bolometers causing spikes in the output voltage timelines.  While these glitches are a nuisance, the effects for bolometric detectors are not as severe as in the case of low-background photoconductive detectors, which are prone to non-linear and memory effects as a result of charge build-up in the semiconductor material. Photoconductors can have operating currents of hundreds of electrons/s or less, but the much higher bias current of a typical cryogenic bolometer ($\sim$ 1 nA or $\sim$ 10$^{9}$ electrons/s) results in the deposited charge being swept out very rapidly, so that after the initial spike has decayed (on a timescale determined by the bolometer thermal time constant) the bolometer has no memory of the event, and its quiescent properties (responsivity and noise) are unaffected by its history of exposure to ionising radiation.

Two types of glitches are seen in the SPIRE detector timelines: large events associated with single direct hits on individual detectors, and smaller co-occurring glitches, seen simultaneously on many detectors in a given array,  that are believed to be due to ionising hits on the silicon substrate that supports all the detectors in an array (frame hits).  The glitch decay time constant is dictated by the bolometer thermal time constant (typically 6 ms), but the on-board analogue signal chains have low-pass electrical filters (designed to eliminate aliased high frequency noise from the sampled timelines), which have the effect of prolonging the duration of a glitch and reducing its amplitude.  Because the photometer channels have a lower cut-off frequency (5 Hz) than the spectrometer channels (25 Hz), spectrometer glitches are detected more easily above the detector noise and so have a higher rate.  With the currently adopted glitch recognition scheme, the detected rates per photometer detector are (0.6, 0.7, 1.5) glitches/detector/minute for (250, 350, 500) $\mu$m, leading to a fractional data loss of less than 1\% in all cases. For the FTS, the equivalent numbers are $\sim$ 3 and 2/detector/minute/ for SSW and SLW respectively.  At present, glitches are identified and flagged in the automatic processing pipelines.  The corresponding data samples are rejected in the photometer map-making stage. In the FTS pipeline, the flagged sections of data are replaced by linear interpolation to  ensure that artefacts are not introduced by subsequent Fourier processing in the pipeline.  Work continues to improve the deglitching in both pipelines, but already glitches are not imposing a significant sensitivity penalty.

\subsection{Photometer performance}
Beam profiles:  The photometer  beam profiles have been measured using scan maps of Neptune, which provides high S/N and appears point-like in all bands (angular diameter $\approx$ 2\arcsec).  The (250, 350, 500)\, $\mu$m beams are well described by 2-D Gaussians down to $\sim$ 15 dB, with mean FWHM values of (18.1, 25.2, 36.6)\arcsec\ and mean ellipticities of (7\%, 12\%, 9\%).  These values, which are very similar to pre-launch predictions, represent an average over each array; there are small systematic variations at the level of $\sim$ 5\% across the arrays.  Beam maps are available (via the ESA \textit{Herschel} Science Centre) allowing the detailed beam shape to be deconvolved from map data if desired.  Each beam map constitutes an averaging in the map over all of the individual bolometers crossing the source, and represents the realistic point source response function of the system, including all scanning artefacts or astrometric uncertainties.  

Flux calibration:  The primary calibrator for the photometer is Neptune, for which we adopt the model of Moreno (1998, 2010), which has an estimated absolute uncertainty of 5\%.  The calibration scheme is defined in detail in Swinyard et al. (2010), Griffin (2010), and in the SPIRE Observers' Manual (2010). Neptune was not observable during most of the PV phase, and an initial calibration was established based on observations of the asteroid Ceres.  At the time of writing, the Neptune-based calibration has yet to be incorporated into the automatic data pipeline, but interim cross-checks have been made to assess the accuracy of the Ceres-based and Neptune-based calibrations, and the overall total uncertainty of the current calibration is estimated as within 15\%.  Further improvement on this figure is expected to be made once the Neptune-based calibration is implemented and through incremental refinements during the course of the mission. Flux densities are quoted at wavelengths of 250, 350 and 500 $\mu$m, based on the \textit{Herschel} convention of  an assumed source SED with flat $\nu$S($\nu$) (i.e., spectral  index -1).  Colour correction factors to convert to a different spectral index are small (a few percent) and within the current 15\% overall uncertainty. 

Sensitivity:  The photometer sensitivity has been estimated from repeated scan maps of dark regions of extragalactic sky. A single map repeat is constituted by two orthogonal scans as implemented in the SPIRE-only scan-map AOT.  Multiple repeats produce a map dominated by the fixed-pattern sky confusion noise, with the instrument noise having integrated down to a negligible value.  This sky map can then be subtracted from individual repeats to estimate the instrument noise.   The extragalactic confusion noise levels for SPIRE are assessed in detail by Nguyen et al. (2010), who define confusion noise as the standard deviation of the flux density in the map in the limit of zero instrument noise.  Measured confusion noise levels in the (250, 350, 500) $\mu$m bands are (5.8, 6.3, 6.8) mJy.  For the nominal scan speed of 30\arcsec/s, the instrument noise is estimated at (9.0, 7.5, 10.8) mJy/sqrt($N_{reps}$) where $N_{reps}$ is the number of map repeats.  These values are comparable to pre-launch estimates at 250 and 500 $\mu$m, and better by a factor of $\sim$ 1.7 at 350 $\mu$m.  For the fast scan speed, the instrument noise levels correspond to (12.7, 10.6, 15.3) mJy/sqrt($N_{reps}$), precisely as expected given the factor of two reduction in integration time per repeat.  For the nominal scan speed, the overall noise is within a factor of sqrt(2) of the (250, 350, 500) $\mu$m confusion levels for (3, 2, 2) repeats.

An important aspect of the photometer noise performance is the knee frequency that characterises the 1/\textit{f} noise of the detector channels.  Pre-launch, a requirement of 100 mHz with a goal of 30 mHz had been specified.  In flight, the major contributor to low frequency noise is temperature drift of the $^{3}$He cooler.  Active control of this temperature is available via a heater-thermometer PID control system, but has not yet been used in standard AOT operation.  The scan-map pipeline (see Sect. 7) includes a temperature drift correction using thermometers, located on each of the arrays, which are not sensitive to the sky signal but track the thermal drifts.  This correction works well and will be improved with a forthcoming update of the flux calibration parameters.  Use of the thermometer signals to de-correlate thermal drifts in the detector timelines over a complete observation (E. Pascale, priv. comm.) can produce a 1/$f$ knee of as low as a 1-3 mHz.  This corresponds to a spatial scale of several degrees at the nominal scan speed.  

Observing overheads: Scan-map mode is very efficient for large area observations, but somewhat less so for smaller fields due to the time needed to turn the telescope around at the end of each scan leg.  The HSpot observation preparation tool (version 4.4.4 at the time of writing) provides a detailed summary of the on-source integration time and the various overheads for any particular observation.  For example, a  2$^\circ$ x 2$^\circ$ map with two repeats takes a total of 7.4 hrs (including all instrument and telescope overheads) with an the overall efficiency of 86\%.  For a 20\arcmin\  x 20\arcmin\  map with two repeats, the duration is 31 min. with an overall efficiency of 51\% due to the larger fraction of time taken up by telescope turn-arounds.  For the Small Map AOT ($\sim$ 5\arcmin\  x 5\arcmin\  map area), with two repeats, the duration is $\sim$ 8 min. with an efficiency of 15\% (but the total observing time still has a large relative contribution from the 3-min. telescope slewing overhead).

The photometer mapping AOTs are designed to ensure that the area requested by the observer is observed with full sampling and highly uniform coverage at the nominal scan speeds.  Maps also contain a region around the target area corresponding to the telescope turn-around periods, in which data are available to the user, but with a lower than nominal scan speed and with less complete spatial sampling.

\subsection{Spectrometer performance}
Wavelength coverage and calibration: The wavelength coverage of the spectrometer is as described in Sect. 4, with the short wavelength SSW band covering 32.0-51.5 cm$^{-1}$ (194-313 $\mu$m) and the long wavelength SLW band covering 14.9-33.0 cm$^{-1}$ (303-671 $\mu$m). Wavelength calibration, verified using CO lines in galactic sources, is accurate within 1/10 of a high-resolution spectral resolution element across both bands.

Beam profiles:   The beamwidths vary across the FTS bands, and this has been characterised using spectral mapping of Neptune as a suitable strong point source. The FWHM varies between 17\arcsec\  and 21\arcsec\  between the short- and long-wavelength ends of the SSW band.  The SLW FWHM exhibits a dip within the band, with values of 37\arcsec\  at the short-wavelength end, 42\arcsec\  at the long-wavelength end, and a minimum of 29\arcsec\  at $\sim$ 425 $\mu$m. More details are given in Swinyard et al. (2010).  

Flux calibration: The spectrometer calibration scheme and methods are described in detail by Swinyard et al. (2010).  At the time of writing, the FTS flux calibration is still based on the asteroid Vesta, but will be updated in the near future based on observations of Neptune and Uranus using the planetary atmosphere models of Moreno (1998, 2010). Current overall calibration accuracy is estimated at 15-30\%, depending on position in the bands, for frequencies above 20 cm$^{-1}$, and will be improved when the planet-based calibration is implemented. Correct subtraction of the thermal emission from the telescope (and from the $\sim$ 5 K instrument enclosure at the lower frequencies) currently requires expert interactive analysis, especially for faint sources, and work continues to establish an automatic pipeline calibration.  The current FTS calibration is adequate for point or extended sources, but for sources partly extended with respect to the beam, calibration is difficult because it requires some a priori knowledge or modelling of the source intensity distribution.  

Sensitivity: The line sensitivity (high-resolution mode; 5~$\sigma$; 1 hr) of the FTS, achieved to date, is typically 1.5 x 10$^{-17}$ W m$^{-2}$ for the SSW band and 2.0 x 10$^{-17}$ W m$^{-2}$ for the SLW band. The sensitivity varies slightly across the band in the manner expected from the shape of the Relative Spectral Responsivity Function (RSRF), as given in the SPIRE Observer's Manual (2010).  At the time of writing, the noise level continues to integrate down as expected for at least 20 repeats (20 forward and 20 reverse scans; total on-source integration time of $\sim$ 45 min.).  The FTS sensitivities are better than pre-launch (HSpot) predictions by a factor of 1.5 -- 2.  This is attributable to the low telescope background, the fact that the SCal source is not used, and to a conservatism factor that was applied to the modelled sensitivities to account for various uncertainties in the model.  

Observing overheads: The telescope pointing is static for most of the time during typical FTS observations, so the overheads are very low. For example, a single pointing high-resolution observation with 20 repeats takes approximately one hour and has an overall efficiency of 86\%.  A fully sampled spatial map, requiring 16 jiggle positions of the BSM, with four repeats per position, takes about 2.4 hrs and is 91\% efficient.

\section{Data-processing pipelines and data quality}
The architecture and operation of the photometer pipeline is described by Griffin et al. (2008), and the spectrometer pipeline is described by Fulton et al. (2008). 

Photometer data are processed in a fully automatic manner, producing measured flux density timelines (Level-1 products) and  maps (Level-2 products).   The pipeline includes conversion of telemetry packets into data timelines, calculation of the bolometer voltages from the raw telemetry, association of a sky position for each detector sample, glitch identification and flagging, corrections for various effects including time constants associated with the detectors and electronics, conversion from detector voltage to flux density, and correction for detector temperature drifts.  The Level-1 products are calibrated timelines suitable for map-making. Maps can be made either with the maximum likelihood map-making algorithm MADmap (Cantalupo et al. 2010) or by "na\"ive" mapping, involving simple binning of the measured flux densities. The pipeline can also be run in an interactive mode, with selectable parameters associated with deglitching, baseline removal and map pixel size.  Standard map pixel sizes of (6, 10, 14)\arcsec\  are adopted for the (250, 350, 500) $\mu$m bands. The most important aspect of the Level-1 data quality that must be addressed prior to map-making is the amount of residual thermal baseline drift on the timelines.  At the time of writing, the standard data products are na\"ive maps generated with pre-treatment of the Level-1 timelines to remove these residual drifts.  The results are high-quality maps, with a low level of scan-related artefacts and with sky structures preserved over large spatial scales (degrees). Work continues to improve further the map-making process, in particular to automate the Level-1 timeline treatment step prior to creation of the maps.  Once that has been completed, the potential advantages of MADmap in producing further enhancements will be investigated.

The FTS pipeline processes the telemetry data producing calibrated spectra. It shares some elements with the photometer pipeline, including the conversion of telemetry into data timelines and the calculation of bolometer voltages. Steps unique to the spectrometer are: temporal and spatial interpolation of the stage mechanism and detector data to create interferograms, apodisation, Fourier transformation, and creation of a hyperspectral data cube. Corrections are made for various instrumental effects including first- and second-level glitch identification and removal, interferogram baseline correction, temporal and spatial phase correction, non-linear response of the bolometers, variation of instrument performance across the focal plane arrays, and variation of spectral efficiency. 

\section{Conclusions}
The SPIRE instrument is fully functional with performance and scientific capabilities matching or exceeding pre-launch estimates. Flux calibration is currently accurate to 15\% for the photometer and 15-30\% for the FTS, and the current pipelines are already producing high-quality data.  Major changes to the AOT implementation are not needed or planned, but further enhancements to the calibration and pipelines will continue as the mission progresses.

\begin{acknowledgements}
SPIRE has been developed by a consortium of institutes led by Cardiff Univ. (UK) and including Univ. Lethbridge (Canada); NAOC (China); CEA, LAM (France); IFSI, Univ. Padua (Italy); IAC (Spain); Stockholm Observatory (Sweden); Imperial College London, RAL, UCL-MSSL, UKATC, Univ. Sussex (UK); Caltech, JPL, NHSC, Univ. Colorado (USA). This development has been supported by national funding agencies: CSA (Canada); NAOC (China); CEA, CNES, CNRS (France); ASI (Italy); MCINN (Spain); SNSB (Sweden); STFC (UK); and NASA (USA).

\end{acknowledgements}


\begin{thebibliography}{}

\bibitem[2006]{ADE} Ade, P. A. R., Pisano, G., Tucker, C. E, and Weaver, S. O. 2006, Proc. SPIE 6275, 62750U

\bibitem[2007]{AUD} Audley, D., Holland, W. S., Atkinson, D., et al. 2007, Proc. ``Exploring the Cosmic Frontier: Astrophysical Instruments for the 21st Century'', Springer-Verlag, p.45

\bibitem[2010]{CAN} Cantalupo, C., Borrill, J. D., Jaffe, A. H., et al. 2010, Ap. J. Suppl., 187, 212

\bibitem[2003]{CHA} Chattopadhyay, G., Bock, J. J., Rownd, B., and Griffin, M. J. 2003, IEEE. Trans. Microwave Theory and Techniques, 51, 2139

\bibitem[2003]{DOL} Dohlen, K. Orign\'{e}, A., Pouliquen, D., and Swinyard, B. 2000, Proc. SPIE 4013, 119, 2000

\bibitem[2003]{DOW} Dowell, C. D., Allen, C. A., Bebu, R., et al. 2003, Proc. SPIE 4855, 73
  
\bibitem[2008]{DUB} Duband, L., Clerc, L., Ercolani, E., et al. 2008, Cryogenics, 48, 95

\bibitem[2002]{FIS} Fischer, J., Klassen, T., Hovenier, N., et al. 2004, \ao, 43, 3765

\bibitem[2008]{FUL} Fulton, T., Naylor, D. A., Baluteau, J.-P., et al. 2008, Proc. SPIE, Vol. 7010, 70102T 

\bibitem[2010]{GRI1} Griffin, M. J., 2010, SPIRE Photometer Flux Density Calibration, SPIRE-UCF-DOC-3168

\bibitem[2002]{GBG} Griffin, M. J., Bock,  J.J., and Gear, W.K. 2002, \ao, 31, 6543

\bibitem[2008]{GRI2} Griffin, M. J., Dowell, C. D., Lim T., et al. 2008, Proc. SPIE, Vol. 7010, 70102Q

\bibitem[2003]{HAR} Hargrave, P. C., Beeman, J. W., Collins, P. A., at al. 2003, Proc. SPIE 4850, 638

\bibitem[2006]{HAR1} Hargrave, P. C., Bock, J. J., Brockley-Blatt, C., et al. 2006, Proc. SPIE 6275, 627513

\bibitem[1998]{MOR1} Moreno, R. 1998, {\it Th\`ese de Doctorat}, Universit\'e de Paris VI

\bibitem[2010]{MOR2} Moreno, R. 2010, Neptune and Uranus planetary brightness temperature tabulation, available fron ESA \textit{Herschel} Science Centre.

\bibitem[2000]{MOS} Moseley, H., Dowell, C. D., Allen, C., et al. 2000, ASP Conference Proceedings, Vol. 217, eds J. G. Mangum and S. J. Radford, 140

\bibitem[2010]{NGU} Nguyen, H., Schulz, B., Levenson, L., et al. 2010, \aap, in press

\bibitem[2010]{PIL} Pilbratt, G. L., Riedinger, J. R., Passvogel, T., et al. 2010, \aap, in press 

\bibitem[2005]{PIS} Pisano, G., Hargrave, P., Griffin, M. J., et al. 2005, \ao  IP, 44, 3208

\bibitem[2010]{POG} Poglitsch, A., Walekens, C., Geis, N., et al. 2010, \aap, in press

\bibitem[2010]{SOM} SPIRE Observers' Manual, 2010, HERSCHEL-HSC-DOC-0789, available from ESA \textit{Herschel} Science Centre

\bibitem[2003]{SW1} Swinyard, B., Dohlen, K., Ferrand, D., et al. 2003, Proc. SPIE 4850, 698 

\bibitem[2003]{SW2} Swinyard, B. M., Ade, P. A. R., Baluteau, J.-P., et al. 2010, \aap, in press

\bibitem[2001]{TUR} Turner, A. D., Bock, J. J., Nguyen, H. T., et al. 2001, \ao 40, 4921 

\end{thebibliography}
\end{document}